\documentclass[12pt,preprint]{aastex}

\slugcomment{Submitted to ApJ}

\shorttitle{Standing Accretion Shocks in the Supernova Core}
\shortauthors{YAMASAKI \& YAMADA}

\begin{document}

\title{Standing Accretion Shocks in the Supernova Core: Effects of Convection and Realistic EOS}

\author{Tatsuya Yamasaki\altaffilmark{1} and Shoichi Yamada\altaffilmark{2,3}}

\altaffiltext{1}{Yukawa Institute for Theoretical Physics,
Kyoto University, Oiwake-cho, Sakyo, Kyoto 606-8502, Japan;
ytatsuya@yukawa.kyoto-u.ac.jp}

\altaffiltext{2}{Science and Engineering, Waseda University, 3-4-1 Okubo,
Shinjuku, Tokyo 169-8555, Japan;
shoichi@heap.phys.waseda.ac.jp}

\altaffiltext{3}{Advanced Research Institute for Science and Engineering,
Waseda University, 3-4-1 Okubo, Shinjuku, Tokyo 169-8555, Japan}

\begin{abstract}

This is a sequel to the previous paper, in which we investigated
the structure and stability of the spherically symmetric accretion flows
through the standing shock wave onto the proto-neutron star
in the post-bounce phase of the collapse-driven supernova.
Following the prescription in the previous paper,
we assume that the accretion flow is in a steady state controlled by the  
neutrino luminosity and mass accretion rate that are kept constant.
We obtain steady solutions
for a wide range of neutrino luminosity and mass accretion rate.
In so doing, as an extension to the previous models,
we employ a realistic EOS and neutrino-heating rates.
More importantly, we take into account
the effect of convection phenomenologically.
For each mass accretion rate, we find the critical neutrino luminosity,
above which there exists no steady solution.
These critical points are supposed to mark the onset of the shock revival. 
As the neutrino luminosity increases for a given mass accretion rate,
there appears a convectively unstable region at some point
before the critical value is reached.
We introduce a phenomenological energy flux by convection
so that the negative entropy gradient should be canceled out.
We find that the convection lowers the critical neutrino luminosity
substantially, which is in accord with the results of multi-dimensional
numerical simulations done over the years.
We also consider the effect of the self-gravity,
which was neglected in the previous paper.
It is found that the self-gravity is important
only when the neutrino luminosity is high.
The critical luminosity, however, is little affected
if the energy transport by convection is taken into account.

\end{abstract}

\keywords{stars: ---
shock waves ---
hydrodynamics ---
supernovae: general}

\section{Introduction}

Core-collapse supernovae are triggered by the gravitational collapse
of massive stars.
The mechanism of explosion is still not well understood.
Many researchers think that the so-called neutrino heating mechanism
is the most promising at present.
In this scenario, after the shock is launched by the core bounce,
it stalls inside the core due to the energy loss
by photo-dissociations of nuclei as well as neutrino-cooling,
and then it is revived by the irradiation of neutrinos 
diffusing out of the proto-neutron star
\citep{wil85}. 
Although many researchers have studied this scenario,
numerical simulations have shown that the stalled shocks do not revive
as long as the collapse is spherically symmetric
\citep{lie01,bur03,tho03,lie05,sum05}.

After the shock is stagnated due to the energy losses, 
the accretion through the standing shock wave
onto the proto-neutron star is approximately steady.
\citet{bur93} and \citet{yam05} attempted to approximate
this phase of evolution by solutions of the time-independent Euler equations,
assuming constant mass-accretion rates and neutrino luminosities.
Adopting such a method, they demonstrated clearly
what is the cause of the failure of explosions found
in the numerical simulations.
Varying these controlling parameters,
the mass-accretion rate and neutrino luminosity, 
they found that for a given mass-accretion rate
there is a critical neutrino luminosity,
above which there exists no steady solution.
Based on this fact, they argued that the revival of stalled shock occurs
when the neutrino luminosity exceeds this critical value.

\citet{yam05} found further that there are in general two types of solutions
for a given mass accretion rate and a neutrino luminosity.
The shock radii in these two solutions differ.
They referred to the solution with a smaller shock radius
as the inner solution and to the other solution as the outer solution.
As the neutrino luminosity is increased with the mass accretion rate fixed,
the shock radius for the inner solution becomes larger,
whereas that for the outer solution becomes smaller.
Two solutions coincide with each other
when the luminosity reaches the critical value.
For the luminosity over the critical value, there is no solution.
They also studied the linear stability of the system
against radial perturbations and found that the inner solution is stable
while the outer solution is unstable and they are merged
at the critical luminosity and become marginally stable.
In that paper, the equation of state and the reaction rates 
for neutrino processes were simplified,
and the self-gravity of the accreting matter is neglected. 
We improve these aspects in this paper
and discuss their effects on the results.

The other concern in this paper is the convection in the accreting flow.
The convection in the supernova core and its implication
for the explosion has been studied by many researchers
using multi-dimensional numerical simulations over the years
\citep{her94,bur95,jan96,mez98}.
They have demonstrated that the convection tends to assist the revival of shock
owing to the non-radial motions and the enhanced heating of the accreting matter
although it seems that the convection alone can not give a successful explosion. 
Our main concern here is whether we can understand
in the frame work by \citet{yam05} the results obtained
by the large scale simulations .
In fact, we have recognized that some solutions
in the paper \citep{yam05} have a region that has a negative entropy-gradient
and is unstable against convective motions.
In this paper, we discuss the effect of convection
by incorporating a phenomenological energy flux
in our time-independent, spherically symmetric models.
The convective energy flux is introduced in such a way that
the negative entropy-gradient should be canceled out.
This assumption will correspond to the maximal mixing by convection and,
hence, we can hopefully estimate the upper limit
for the effect of convection on the critical neutrino luminosity.

\section{Models}

In this section, we describe our models,
giving the basic equations and some assumptions.
We are interested in the post bounce phase,
in which the shock is stalled inside the core
and becomes a standing accretion shock.
Matter is accreted onto the proto-neutron star through the shock wave, 
being irradiated by neutrinos diffusing out of the proto-neutron star. 

Assuming that this accretion flow is steady and spherically symmetric,
we solve the following time-independent ordinary differential equations.
\begin{equation}
4\pi r^2 \rho u_r=\dot{M},
\label{eq1}
\end{equation}
\begin{equation}
u_r \frac{du_r}{dr}+\frac{1}{\rho}\frac{dp}{dr}+\frac{GM}{r^2}=0,
\label{eq2}
\end{equation}
\begin{equation}
u_r \rho T\frac{dS}{dr}
=\dot{q},
\label{eq3}
\end{equation}
\begin{equation}
u_r n \frac{dY_ e}{dr}
=\lambda,
\label{eq4}
\end{equation}
\begin{equation}
\frac{dM}{dr}=4\pi r^2 \rho,
\label{eq5}
\end{equation}
where $r$, $u_r$, $\rho$, $p$, $S$, $Y_e$ and $n$ denote the radius,
radial velocity, density, pressure, entropy per unit mass, electron fraction
and baryon number density, respectively;
$\dot{M}$, $G$ and $M$ are the mass accretion rate, gravitational constant and
mass enclosed in the sphere of radius $r$, respectively;
$\dot{q}$ and $\lambda$ are the heating and reaction rates
due to the reactions with neutrinos, respectively.
The rotation and magnetic field are neglected.
The Newtonian formulation is justified
because the region of our interest is outside the proto-neutron star
and the general relativistic correction is at most 10\%.

We do not consider the flow outside the shock wave, simply assuming a free fall,
and restrict our calculations to the region inside the shock.
We impose the Rankine-Hugoniot relations at the shock front, $r=r_{\rm sh}$:
\begin{equation}
\rho u_r^2 +p=\rho_{\rm u} u_{\rm u}^2 +p_{\rm u},
\label{eq10}
\end{equation}
\begin{equation}
\frac{1}{2} u_r^2 +\epsilon +\frac{p}{\rho}
=\frac{1}{2} u_{\rm u}^2
+\epsilon_{\rm u}+\frac{p_{\rm u}}{\rho_{\rm u}},
\label{eq11}
\end{equation}
where $\epsilon$ denotes the internal energy
including the nuclear binding energy.
The suffix ${\rm u}$ stands for the upstream quantities given as follows.
Assuming that the upstream matter is composed of cold irons
and flowing into the shock at a free-fall velocity, we take 
\begin{equation}
S_{\rm u}=0,
\label{eq12}
\end{equation}
\begin{equation}
Y_{e,{\rm u}}=Y_e=\frac{26}{56},
\label{eq13}
\end{equation}
and
\begin{equation}
u_{\rm u}=\sqrt{\frac{2GM}{r}}.
\label{eq14}
\end{equation}
The inner boundary of the calculation region is set at the neutrino sphere, 
from which, we assume, thermal neutrinos of all flavors with a temperature of
$T_{\nu}=4.5 {\rm MeV}$ are radiated outwards.
The radius of the neutrino-sphere,
$r_{\nu}$, satisfies the following relation,
\begin{equation}
L_{\nu_e}=\frac{7}{16}\sigma T_{\nu}^4 \cdot 4\pi r_{\nu}^2 ,
\label{eq15}
\end{equation}
where $\sigma$ is the Stefan-Boltzmann constant.
At the inner boundary, we also assume  
\begin{equation}
M=1.3M_{\odot}, 
\label{eq16}
\end{equation}
that is, the mass of the proto-neutron star.

As for the neutrino reactions, we take into account only the emission and
absorption by free nucleons.
\begin{equation}
p + e^- \rightleftharpoons \nu_e + n,
\label{eq6}
\end{equation}
\begin{equation}
n + e^+ \rightleftharpoons \bar{\nu}_e + p.
\label{eq7}
\end{equation}
The heating and reaction rates can be decomposed accordingly as
\begin{equation}
\dot{q}=\dot{q}_{\rm ep}-\dot{q}_{\nu{\rm n}}+\dot{q}_{\rm e^+ n}
-\dot{q}_{\bar{\nu}{\rm p}},
\label{eq8}
\end{equation}
\begin{equation}
\lambda=-\lambda_{\rm ep}+\lambda_{\nu{\rm n}}+\lambda_{\rm e^+ n}
-\lambda_{\bar{\nu}{\rm p}}.
\label{eq9}
\end{equation}
These rates are calculated based on the formulae given by \citet{bru85}.
In the previous paper \citep{yam05},
a simplified equation of state was employed
and the photo-dissociations of nuclei are neglected for simplicity.
In this paper, we employ a realistic equation of state by \citet{she98}
and take into account the nuclear composition and dissociation consistently.

The optical depth of the inner boundary (or the neutrino sphere)
from the shock front is set to be $2/3$ for the electron-type neutrino.
\begin{equation}
\tau_{\nu_e}\equiv\int^{r_{\rm sh}}_{r_{\nu}}\alpha_{\nu_e}dr=\frac{2}{3},
\label{eq17}
\end{equation}
where $\alpha_{\nu_e}$ is the energy-averaged absorption coefficient
for the reactions of the electron-type neutrino given in equation (\ref{eq6}).
We neglect the contribution from the layer outside the shock wave
to the optical depth because the absorption coefficient is
negligibly small there.
Since the region of our interest is outside the neutrino sphere, 
we do not solve the neutrino transfer equations, assuming that the luminosity
and energy of neutrinos are independent of radius.

Adopting the neutrino luminosity, $L_{\nu_e}$, and 
the mass accretion rate, $\dot{M}$, as model parameters,
we solve equations (\ref{eq1})-(\ref{eq5})
together with the boundary conditions for a wide range of these parameters.
The phenomenological treatment of convection
will be described in section \ref{sub_conv}.

\section{Results}

\subsection{Effects of Improved Physics\label{sub_imp}}

\begin{figure}
\plottwo{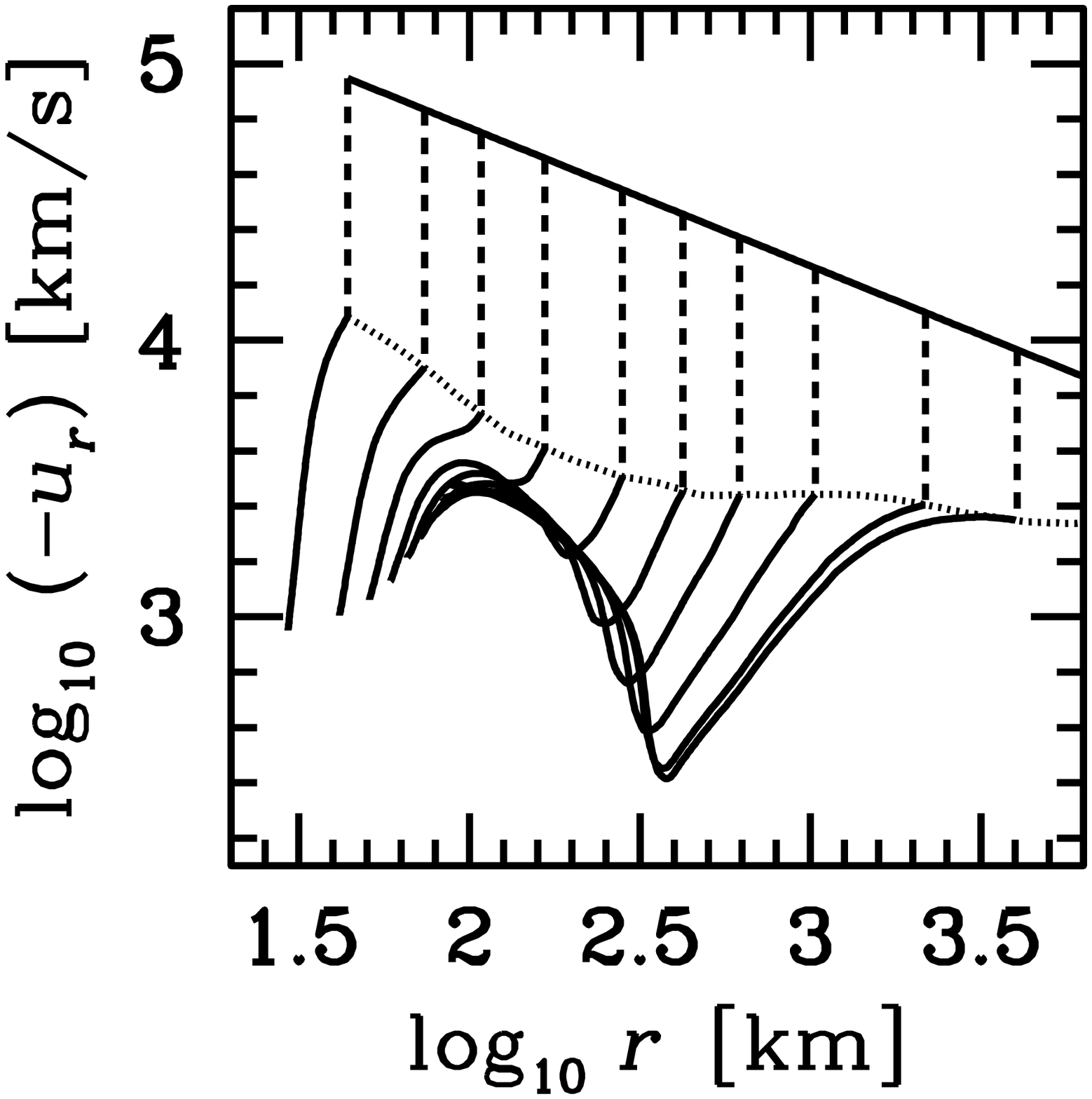}{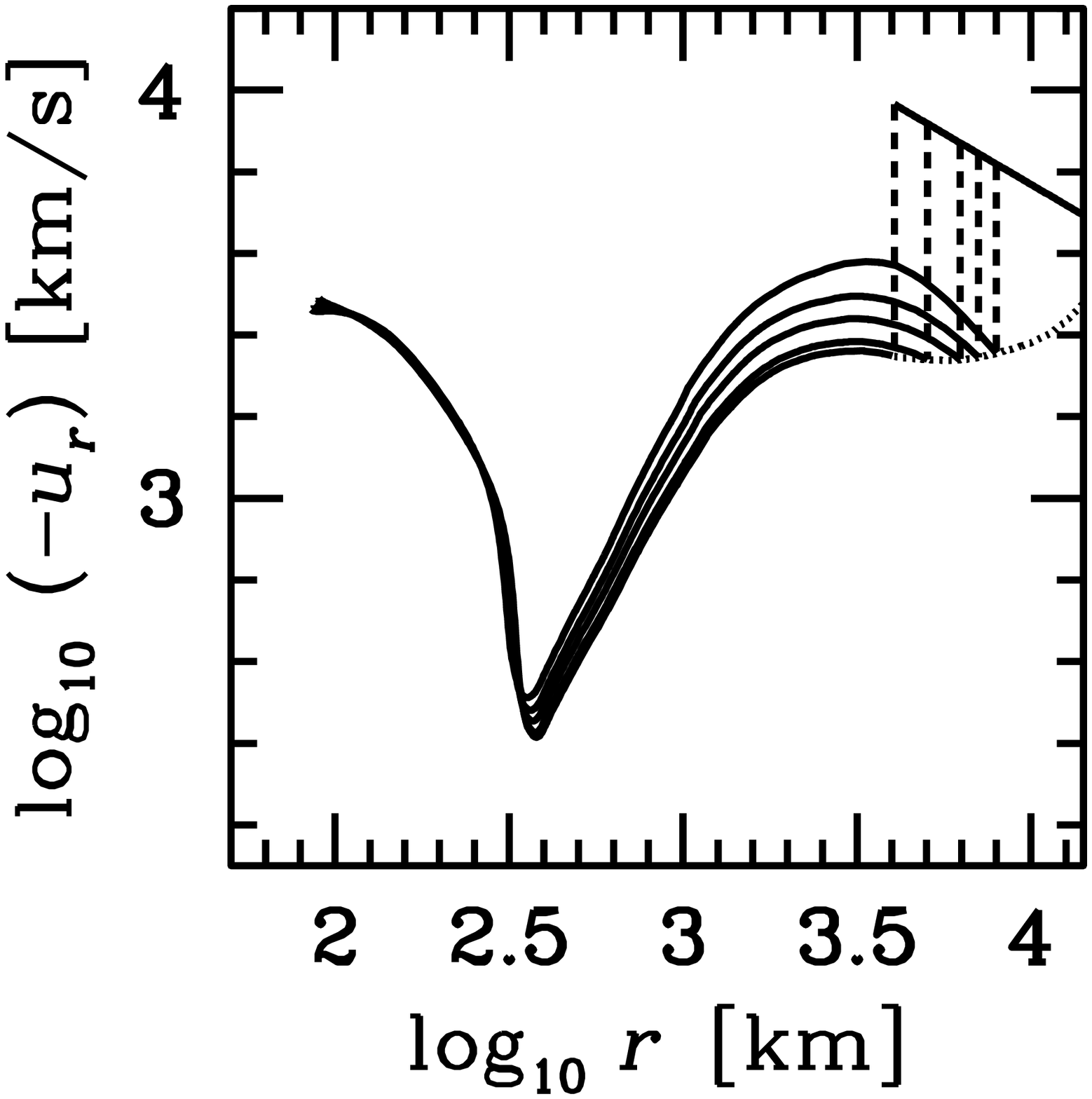}
\caption{Left. Accretion velocity for the inner solutions
with $\dot{M}=1.0$ $M_{\odot}$ ${\rm s}^{-1}$,
$L_{\nu_e}=(2,4,6,8,10,12,14,16,18)\times 10^{52}$ ${\rm ergs}$ ${\rm s}^{-1}$
and $18.7125 \times 10^{52}$ ${\rm ergs}$ ${\rm s}^{-1}$
(critical value) from left to right.
Right. Accretion velocity for the outer solutions
with $\dot{M}=1.0$ $M_{\odot}$ ${\rm s}^{-1}$,
$L_{\nu_e}=(17.0,17.5,18.0,18.5)\times 10^{52}$
${\rm ergs}$ ${\rm s}^{-1}$
and $18.7125 \times 10^{52}$ ${\rm ergs}$ ${\rm s}^{-1}$ from top to bottom.
Dashed lines denote shock jumps. Dotted lines show the downstream velocity
satisfying the Rankine-Hugoniot relations at each radius.
The self-gravity is not taken into account.
\label{fig1}}
\end{figure}

First, we show the results of the calculations with the self-gravity neglected.
We fix the mass to be $M=1.3M_{\odot}$ and ignore equation~(\ref{eq5}).
The radial infall velocities of the spherically symmetric steady flows
are given in figure \ref{fig1}. Although we have obtained solutions for various values of mass accretion rate,
we show only the case with $\dot{M}=1.0{M}_{\odot}{\rm s}^{-1}$.
For other values of $\dot{M}$, the results are similar qualitatively.
As in the previous paper \citep{yam05}, we find that for a given mass
accretion rate and neutrino luminosity,
there are two types of solutions, each having different radii
(that is, the inner solution and outer solution)
as long as the neutrino luminosity is below a certain critical value.
As shown in figure \ref{fig1}, the shock radius for the inner solutions
becomes larger, whereas that for the outer solutions becomes smaller,
as the neutrino luminosity is raised with the mass accretion rate fixed.
These two solutions coincide with each other
when the luminosity reaches the critical value, and there is no solution 
for the luminosity exceeding the critical value.

\begin{figure}
\plotone{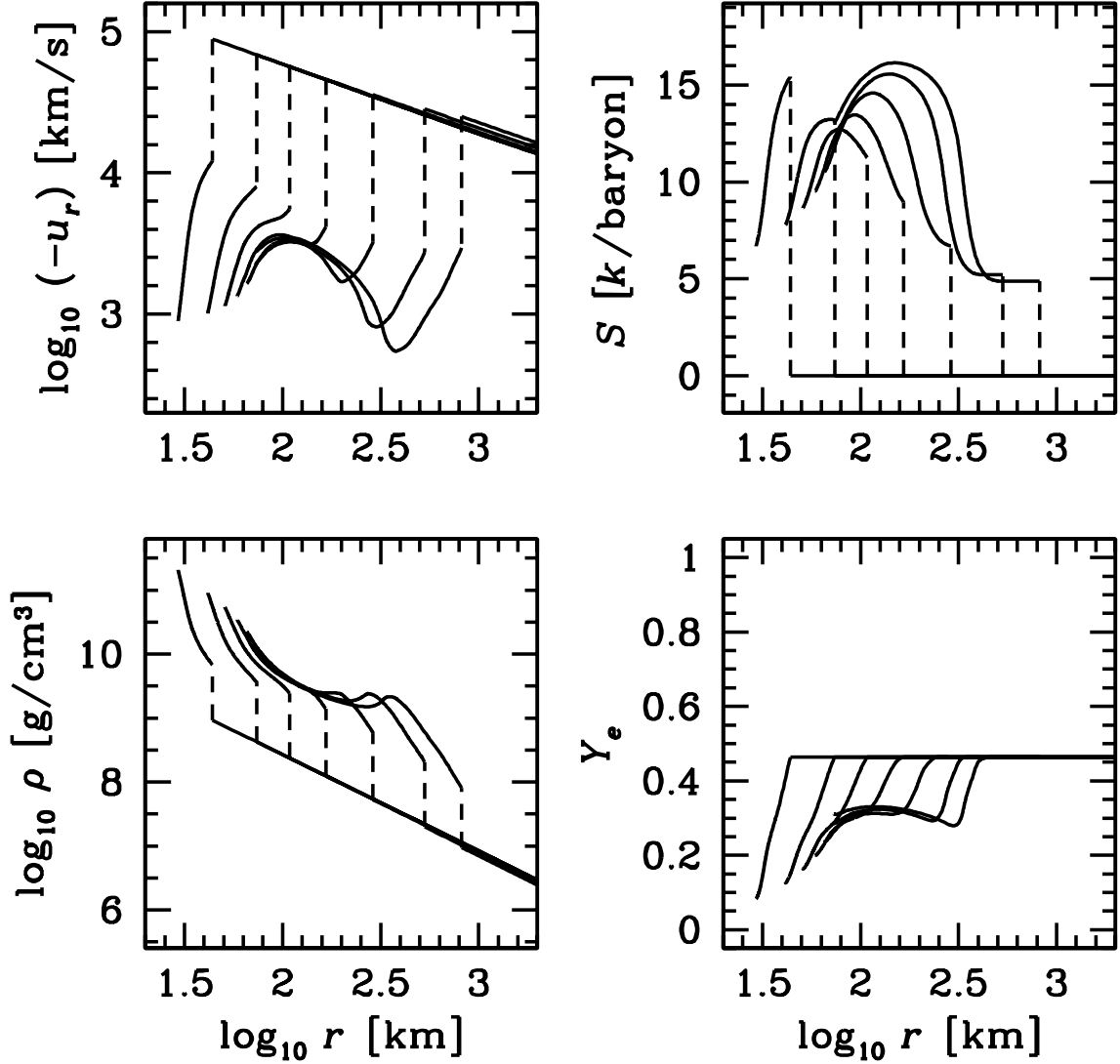}
\caption{Accretion velocity (upper-left panel),
density (lower-left panel), entropy (upper-right panel),
electron fraction (lower-right panel)
for $\dot{M}=1.0$ $M_{\odot}$ ${\rm s}^{-1}$,
$L_{\nu_e}=(2,4,6,8,10,12)\times 10^{52}$ ${\rm ergs}$ ${\rm s}^{-1}$
and $12.4606 \times 10^{52}$ ${\rm ergs}$ ${\rm s}^{-1}$
(critical value) from left to right.
\label{fig2}}
\end{figure}

This behavior of the solutions is understood as follows.
Equations (\ref{eq1})-(\ref{eq4}) can be re-assembled for the radial velocity as,
\begin{equation}
\left\{u_r^2 -\left(\frac{\partial p}{\partial \rho}\right)_{S,Y_e}\right\}
\frac{d\ln |u_r|}{dr}
=\frac{2}{r}\left(\frac{\partial p}{\partial \rho}\right)_{S,Y_e}
-\frac{GM}{r^2}
-\frac{1}{\rho}\left\{\left(\frac{\partial p}{\partial S}\right)_{\rho,Y_e}
\frac{\dot{q}}{u_r \rho T}
+\left(\frac{\partial p}{\partial Y_e}\right)_{\rho,S}
\frac{\lambda}{u_r n}\right\}.
\label{eqa}
\end{equation}
As known for the Bondi flow \citep{sha83}, the subsonically accreting matter is
first accelerated and then decelerated, 
giving a peak in the $r-u_{r}$ diagram
(see the region with $\log_{10}r\sim 3.5$ in figure \ref{fig1}).
[Note that the post shock matter is not hydrostatic
and the advection cannot be ignored particularly
when the shock radius is larger than about $10^8$cm
for the mass accretion rate of 1 ${\rm M}_{\odot}/{\rm s}$.]
This is because the first term on the right-hand side of the equation
(\ref{eqa}) dominates the other terms in the outer region, 
whereas the second term becomes dominant in the inner region.
The third and forth terms on the right-hand side are negligible at large radius.
Since the free-fall velocity decreases monotonically with $r$,
there are in general two solutions satisfying the Rankine-Hugoniot relations. 
When the neutrino luminosity increases, the accretion velocity is lowered, 
and the solution ceases to exist at some point,
which gives the critical neutrino luminosity. 
In the present case, however, the behavior is more complicated.
The accretion velocity has a second peak
(see the region with $\log_{10}r\sim 2$ in figure \ref{fig1})
when the luminosity becomes large and the heating zone appears.
The emergence of the second peak is attributed to the neutrino heating and
cooling represented by the third and forth terms on the right-hand side of
equation (\ref{eqa}). Since $(\partial p/\partial S)_{\rho,Y_{e}}$ is positive
and the last term is always smaller than the third term on the right-hand side
of equation (\ref{eqa}), the non-adiabatic effects represented by the last two
terms work so as to reduce the deceleration of flows in the heating region.
In addition, as the inflowing matter is heated, the value of
$(\partial p/\partial \rho)_{S,Y_{e}}$ becomes larger, and if the heating is
great enough, the sum of the right-hand side of equation (\ref{eqa}) becomes
positive at some point and the flow switches from deceleration to acceleration. 
After matter enters the cooling region, the value of
$(\partial p/\partial \rho)_{S,Y_{e}}$ decreases again and the second term on
the right-hand side of equation (\ref{eqa}) dominates over the other terms 
at some radius and, as a result, the flow is decelerated again in the innermost
region (the upper left panel of figure \ref{fig2}). 
The improvement of the EOS and the proper treatment
of the photo-dissociations of nuclei are responsible for the difference
from the results in the previous paper.
The temperature is substantially smaller in the present models.
Apart from this non-monotonic behavior, however,
the characteristics of the solutions found in the previous paper still hold.
We think that the outer solution will not be realized in the actual evolution
of the supernova core because the shock radius for the outer solution is
extremely large for rather small luminosities that are commonly found by
detailed numerical simulations in the phase of the shock stagnation.
Hence we will discuss only the inner solutions from now on.

\begin{figure}
\plottwo{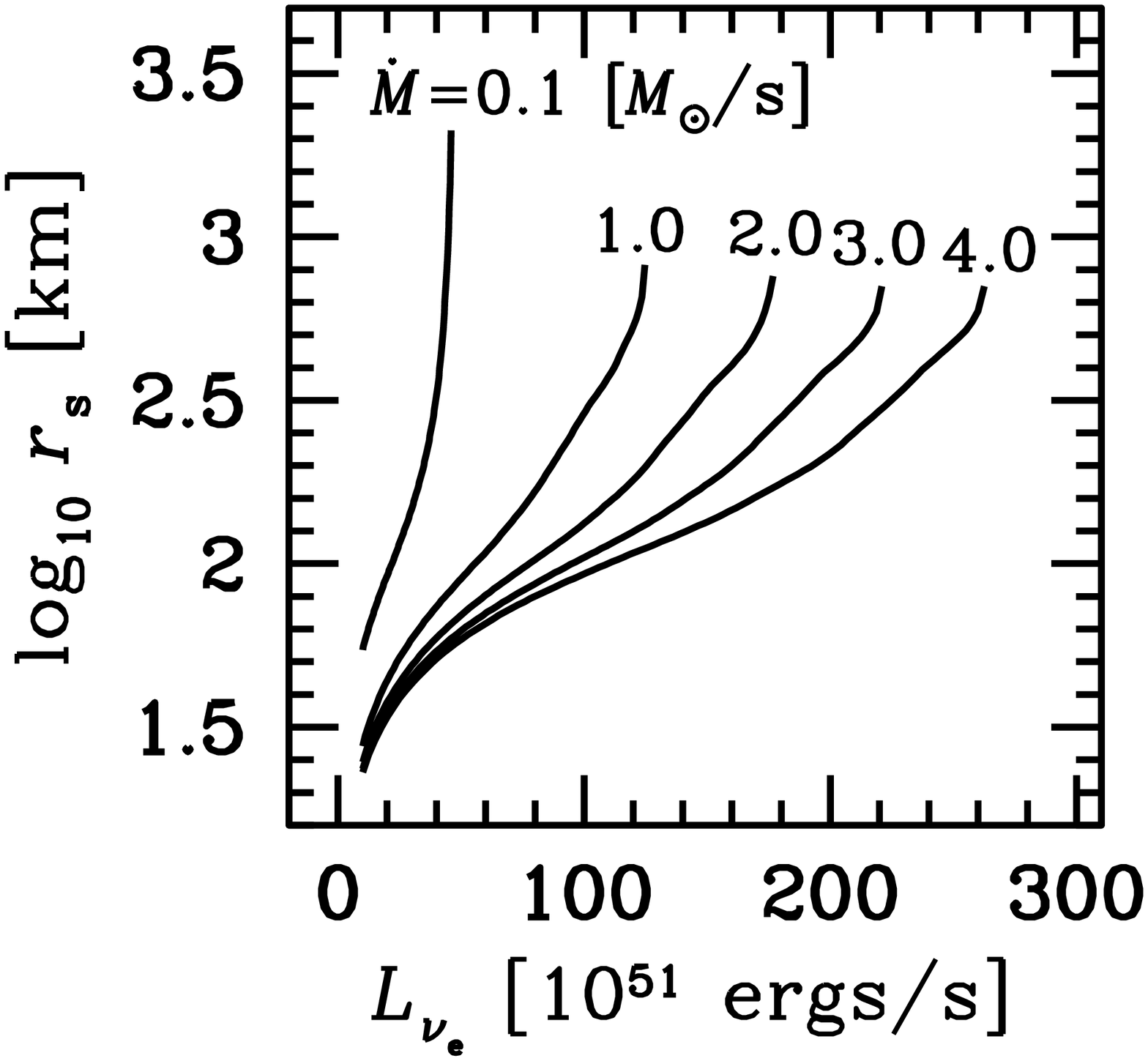}{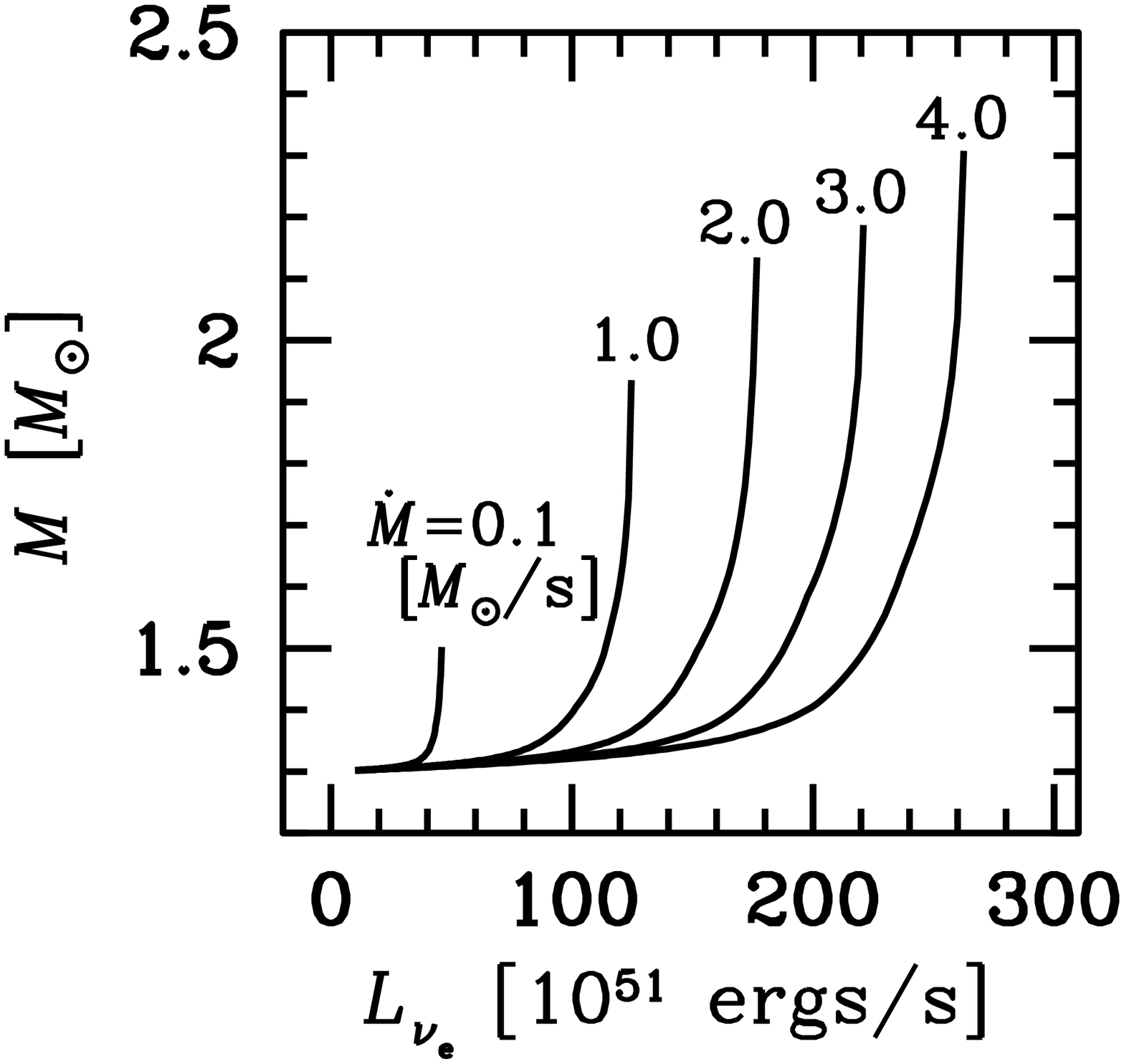}
\caption{Left. Radius of the shock surface
for $\dot{M}=0.1, 1.0, 2.0, 3.0,
4.0$ ${M}_{\odot}$ ${\rm s}^{-1}$ from left to right.
Right. Mass enclosed in the shock surface
for $\dot{M}=0.1, 1.0, 2.0, 3.0,
4.0$ ${M}_{\odot}$ ${\rm s}^{-1}$ from left to right.
\label{fig3}}
\end{figure}

The results with self-gravity are shown in figure \ref{fig2}.
As long as the neutrino luminosity is less than about
$12\times 10^{52}{\rm ergs~s^{-1}}$, the effect of self-gravity is negligible,
and the solution curves are almost the same as
those neglecting the self-gravity,
while the effect becomes significant quickly
when the luminosity exceeds this value.
As can be seen in figure \ref{fig3},
where we show the shock radius and the mass
enclosed in the shock surface, they increase rapidly
as the luminosity approaches the critical value.
The effective increase of gravitational attraction owing to
the incorporation of self-gravity in the second term
on the right-hand side of equation (18) tends to decrease
the velocity of the subsonic flow. This may appear to be counter-intuitive,
but can be understood as follows.
When the accretion flow is supersonic, the pressure is less important
and the increase of mass simply results in the increase of flow speed.
For the subsonic flow, on the other hand,
the pressure plays an important role and the increase of gravity
results in the increase of pressure as well as the density,
which in turn leads to the decrease of the flow velocity
because of the mass conservation.
The effect of self-gravity is negligible for the inner
solutions in most cases, while it seriously affects the outer solutions,
in which the flow is initially accelerated. Since the density is increased,
the shock radius becomes smaller so that the optical depth for neutrinos be 2/3
at the inner boundary (\ref{eq17}). 
This in turn leads to the smaller critical luminosity,
at which the outer solution coincides 
with the inner solution and the steady solution disappears.
It is interesting to see that the shock radius for the critical luminosity
is almost unchanged as the mass accretion rate is varied.
It becomes slightly larger when the mass accretion rate
gets smaller than $1.0 M_{\odot}$.

\begin{figure}
\plottwo{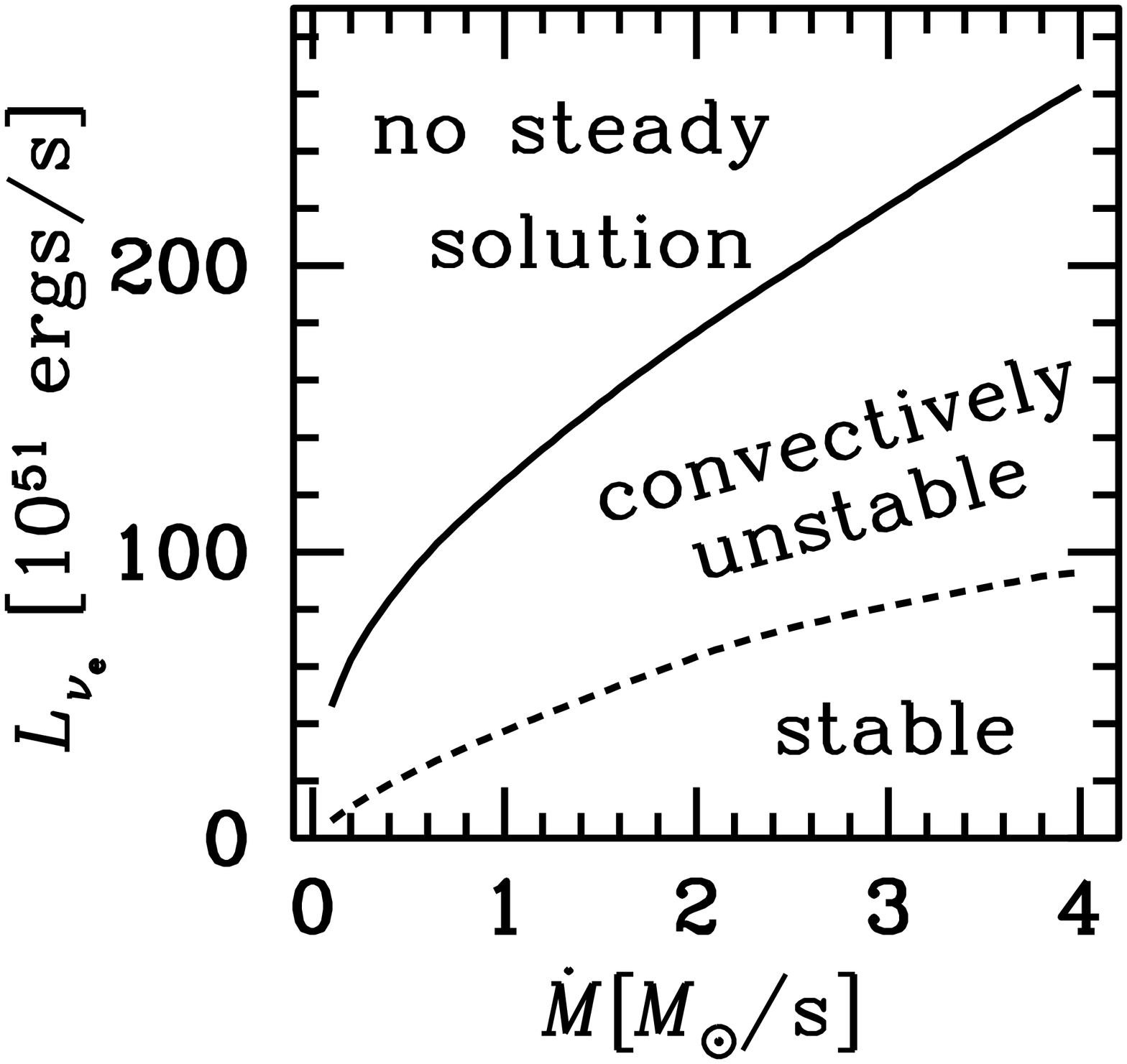}{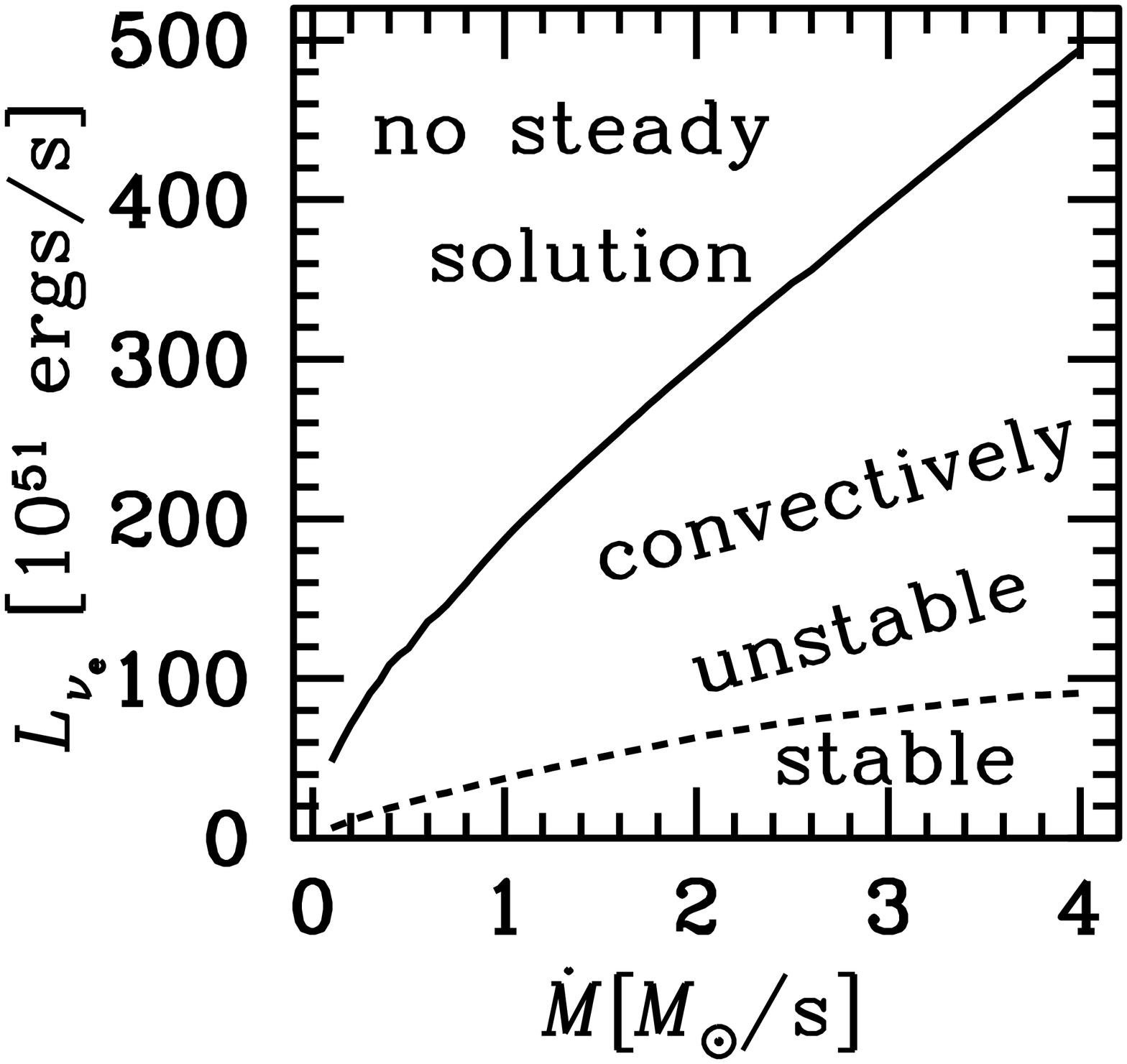}
\caption{Left. Critical luminosity (solid curve) and boundary for 
convective/non-convective solutions (dashed curve). 
Self-gravity is taken into account.
Above the dashed curve, there exists a convectively unstable region in the solution.
Right. The same as in a. but without self-gravity.
\label{fig4}}
\end{figure}

In figure \ref{fig4}, we show the critical luminosities for various
mass accretion rates, both with and without self-gravity taken into account. 
As shown also in figure \ref{fig2}, there emerges a region
with a negative entropy-gradient when the luminosity exceeds a certain value
for each mass accretion rate.
Such a region is convectively unstable by the classical criterion 
(see section 4 for possible corrections to the criterion),
We will discuss the effect of convection in detail
in section \ref{sub_conv}.
Since the self-gravity becomes appreciable quickly once the luminosity exceeds 
a certain value that is slightly smaller than the critical luminosity,
the critical luminosity is significantly affected by the self-gravity
as long as the convection is ignored.
Comparing figure \ref{fig4} with the previous results
(figure 1 of \citet{bur93} or figure 8 of \citet{yam05}),
we can see that the critical luminosity is substantially increased
in the present models.
In section \ref{sub_conv} we will discuss
how this will be changed again after the
convection is taken into account.

\subsection{Effect of Convection\label{sub_conv}}

In subsection \ref{sub_imp}, we have seen that
the accretion flow has in general
a region with a negative entropy-gradient if the neutrino luminosity is
larger than a certain value.
Then, we expect and multi-dimensional numerical simulations have demonstrated
that the convective motions occur.
The convection induces not only the out-going energy flow effectively
but also the mixing of matter, and, as a result,
affects the structure of the accretion flow and the location of the shock.
In this section, we discuss the effect of the convection on the steady flow
as well as on the critical neutrino luminosity in the present frame work.

Recently, \citet{fog05} pointed out that the negative entropy-gradient is 
not a sufficient condition for the convection in the accreting matter and that 
the convection will be suppressed by the advection. 
In section \ref{sub_imp}, we used a classical criterion to judge if there is a 
convection zone or not, which would have been an overestimation
for the convective region.
In this section, we still stick to the classical criterion,
since the approach proposed below is highly phenomenological
and the results we obtain should be regarded as 
"qualitative", but not "quantitative".
The possible suppression of convection due to 
advection in our models will be assessed in section 4.

In order to incorporate the effect of convection
in our spherically symmetric models, 
we introduce the effective energy- and proton-number fluxes as follows.
In the convection zone ($r>r_{\rm gain}$), the equations for energy (\ref{eq3})
and electron fraction (\ref{eq4}) are replaced by
\begin{equation}
u_r \rho T\frac{dS}{dr}
=\dot{q}-\frac{1}{r^2}\frac{dr^2 F_{e,{\rm conv}}}{dr},
\label{eq18}
\end{equation}
\begin{equation}
u_r n \frac{dY_ e}{dr}
=\lambda-\frac{1}{r^2}\frac{dr^2 F_{n,{\rm conv}}}{dr},
\label{eq19}
\end{equation}
where $F_{e,{\rm conv}}$ and $F_{n,{\rm conv}}$ are the energy-
and the proton-number fluxes associated
with the non-spherical convective motions, respectively. 

It is admittedly difficult to give these fluxes from the first principle.
Multi-dimensional numerical simulations done so far suggest that
the angle-averaged treatment like the one employed in this paper
may not be a very good approximation for the convection in the heating region.
With these caveats in mind, however, we proceed as follows.
Recalling that the convection is driven by the negative entropy-gradient
and tends to reduce it, we assume in the convective zone that 
the right-hand sides of equations (\ref{eq18}) and (\ref{eq19}) vanish,
that is,
\begin{equation}
\dot{q}-\frac{1}{r^2}\frac{dr^2 F_{e,{\rm conv}}}{dr}=0,
\label{eq20}
\end{equation}
and 
\begin{equation}
\lambda-\frac{1}{r^2}\frac{dr^2 F_{n,{\rm conv}}}{dr}=0.
\label{eq21}
\end{equation}
This is in a sense the limit of maximally efficient mixing,
which realizes an isentropic distribution in the convective zone.
We regard these equations as the definitions of the fluxes.

Since the convective region is extended up to the shock front,
these fluxes should be also taken into account
in the jump conditions for the shock wave.
We modify equations~(\ref{eq11}) and (\ref{eq13}) in the original condition to
\begin{equation}
\frac{1}{2} u_r^2 +\epsilon +\frac{p}{\rho}
+\frac{F_{e,{\rm conv}}}{\rho u_r}
=\frac{1}{2} u_{\rm u}^2
+\epsilon_{\rm u}+\frac{p_{\rm u}}{\rho_{\rm u}},
\label{eq22}
\end{equation}
\begin{equation}
Y_e +\frac{F_{n,{\rm conv}}}{n u_r}
=Y_{e, {\rm u}}.
\label{eq23}
\end{equation}
We further impose the following conditions, 
\begin{equation}
F_{e,{\rm conv}}=F_{n,{\rm conv}}=0.
\label{eq25}
\end{equation}
at the inner boundary of the convective zone, that is, 
the gain radius $r=r_{\rm gain}$, where 
\begin{equation}
\dot{q}=0,
\label{eq24}
\end{equation}
is satisfied.

\begin{figure}
\plotone{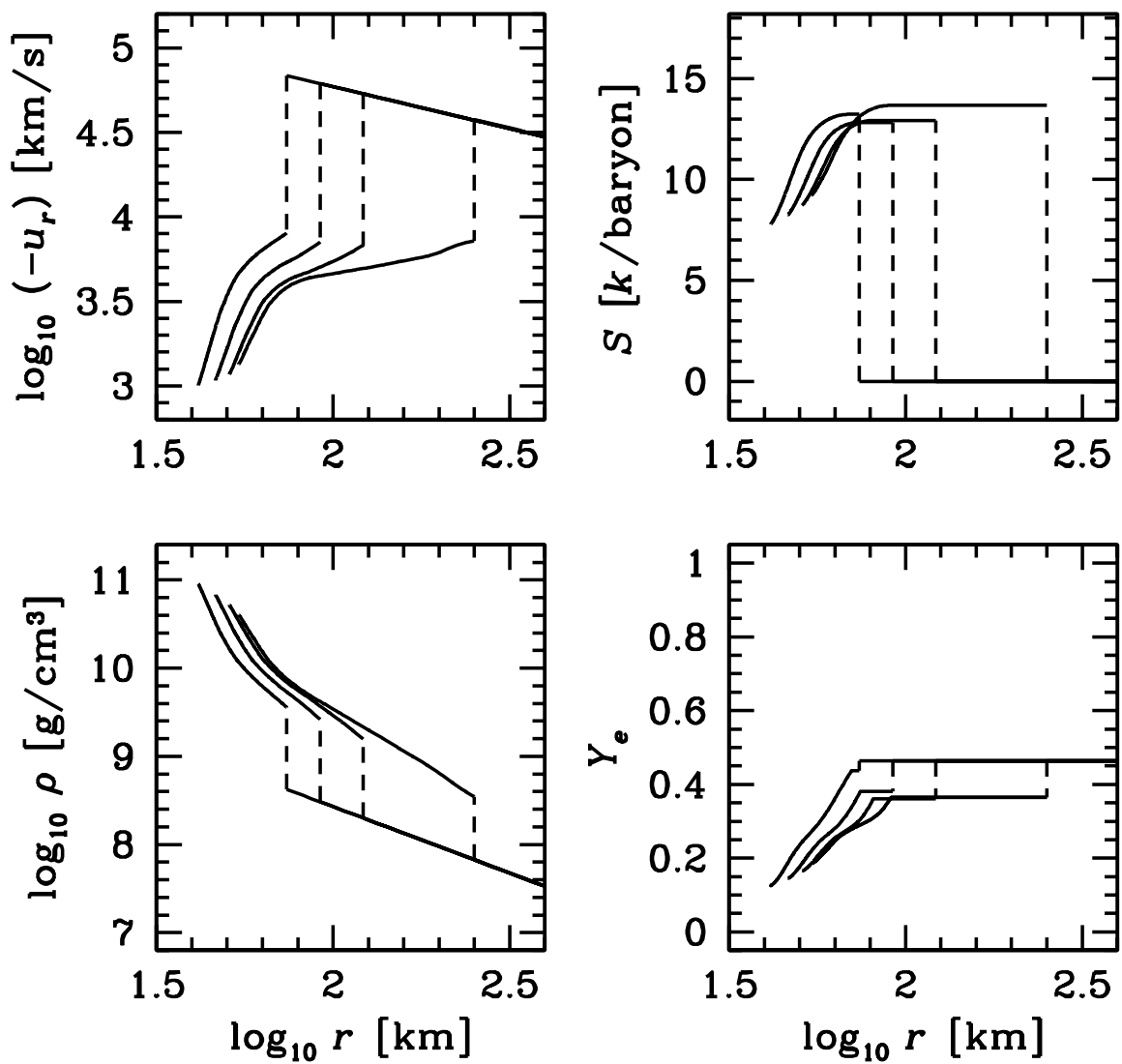}
\caption{As in Fig. 2, but with convection taken into account and 
with $\dot{M}=1.0$ $M_{\odot}$ ${\rm s}^{-1}$,
$L_{\nu_e}=(4,5,6)\times 10^{52}$ ${\rm ergs}$ ${\rm s}^{-1}$
and $6.7860 \times 10^{52}$ ${\rm ergs}$ ${\rm s}^{-1}$
(critical value) from left to right.
Dashed lines denote the shock jumps.
\label{fig5}}
\end{figure}

\begin{figure}
\plotone{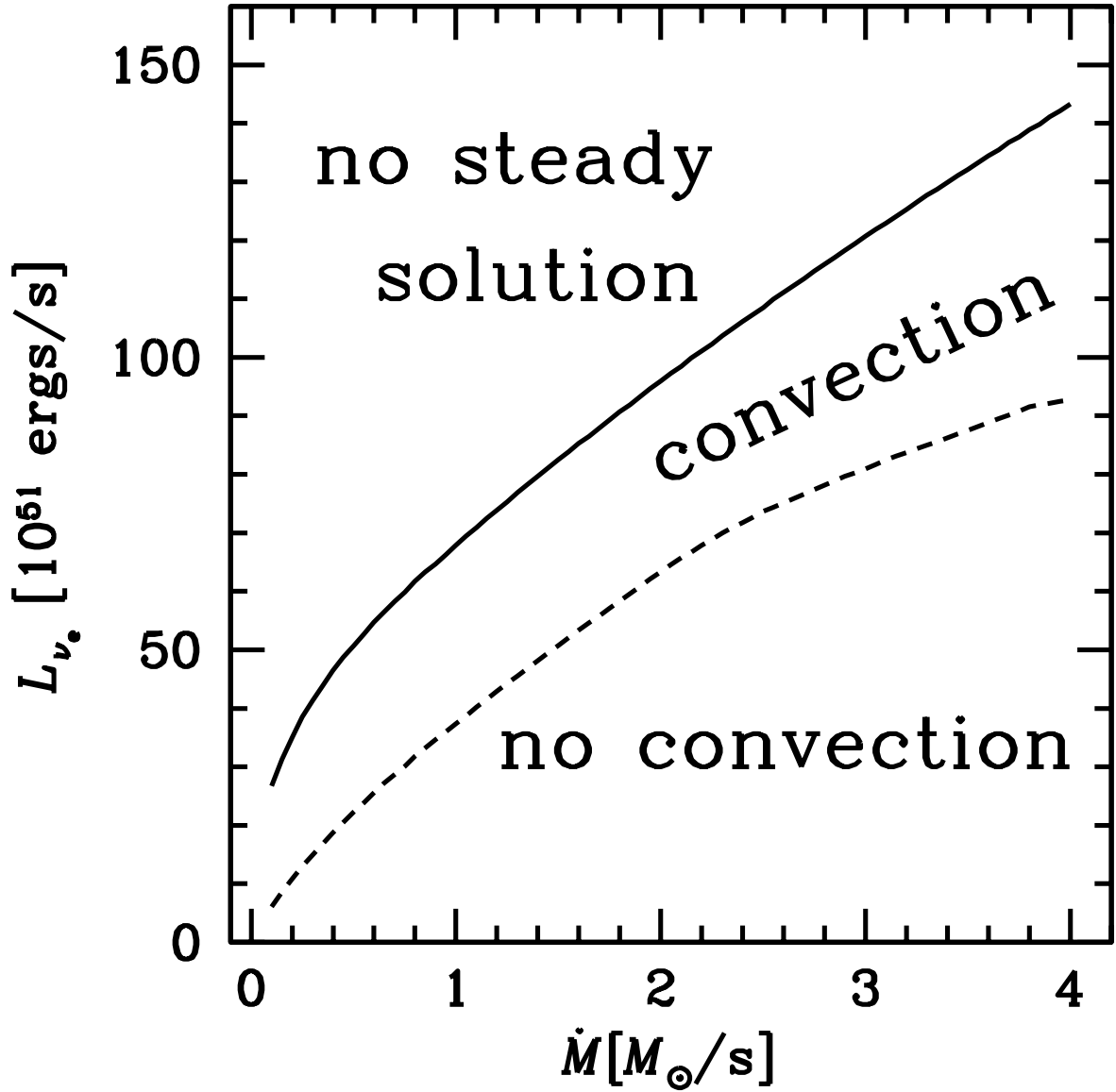}
\caption{Critical luminosity (solid curve) for the models with convection taken into account.
Above the dashed curve, there is a convection zone in the solution.
\label{fig6}}
\end{figure}

\begin{figure}
\plotone{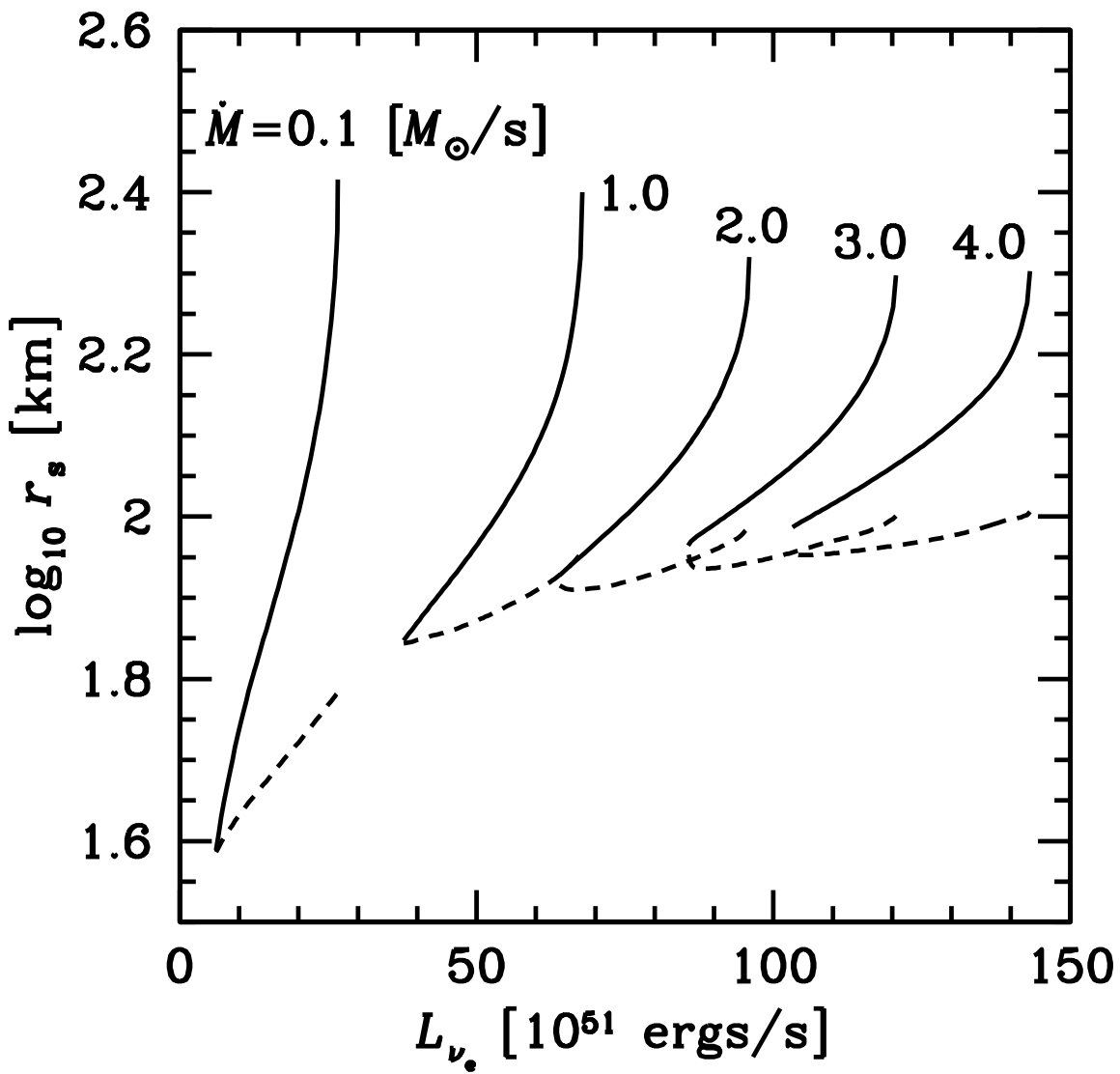}
\caption{Radius of the shock surface (solid curves) for the models with convection 
taken into account.
The mass accretion rates are $\dot{M}=0.1, 1.0, 2.0, 3.0,
4.0$ ${M}_{\odot}$ ${\rm s}^{-1}$ from left to right.
Dashed curves denote the gain radius.
\label{fig7}}
\end{figure}

The solutions of the above equations are shown
in figures \ref{fig5} to \ref{fig7}, 
where the self-gravity is taken into account.
In figure \ref{fig5}, we can see that 
the complicated post-shock velocity distributions with a double-peak have been 
replaced by the monotonic ones after the convection is taken into account.
This is understood as follows. In our treatment of convection,
the last two terms on the right-hand side of equation (\ref{eqa}) is
just compensated by the convective flux in the heating region.
As a result, the situation in the heating region becomes
very much similar to the adiabatic flow, and the flow is monotonic again.
More importantly, the critical luminosity is lowered by a factor of $\sim2$
by the convection as demonstrated in figure \ref{fig6}.
This agrees qualitatively with the results of
multi-dimensional numerical simulations.
In terms of our model, the reason is the following.
The convection tends to raise the entropy just behind the shock wave.
This leads to the decrease in the accretion velocity, which then
makes it easier to reach the critical point.

It is also noted that the shock radius for the critical luminosity
becomes smaller (see figure \ref{fig7}),
compared with the models neglecting the convection.
This implies that the results are hardly affected
by self-gravity if the convection is 
taken into account, since the self-gravity becomes
important only when the critical point
is approached in the models without convection.

\section{Summary and Discussion}

In this paper, we have obtained spherically symmetric steady accretion flows 
through the standing shock wave onto the proto-neutron star.
We have improved the treatment of microphysics such as the equation of state
and neutrino reaction rates.
We have found that our previous findings are essentially
unchanged after these modifications.
For a given mass accretion rate, there are in general two solutions, 
the inner and outer solutions.
There is a critical neutrino luminosity for a given
mass accretion rate, above which no steady solution exists.
The outer solution is 
unstable against radial perturbations and the two branches of solution
coincide with each other for the critical luminosity.
There is a quantitative difference, though.
The critical luminosity, in particular, becomes much higher
when the realistic microphysics are implemented. 

We have also investigated the effect of convection,
employing a phenomenological description of convection.
We have introduced the convective fluxes of energy and proton number,
which then have been determined in such a way that
the negative entropy gradient is canceled out by them.
This is roughly corresponding to the maximally efficient mixing by convection. 
We have found that the critical luminosity is lowered by a factor of $\sim2$
by the convection, consistent with results
from the large-scale numerical simulations done so far.
The latest results of detailed numerical simulations suggest that
the convection is not sufficient for the shock revival \citep{bur06}.
Comparing the values of the critical luminosity in our models
with the actual luminosities obtained by detailed simulations \citep{sum05},
we have found that the former is larger.
Thus, our results agree with the numerical simulations also in this respect.
It is noted that our models are supposed to correspond
to the maximal convection. 
Furthermore, we have found that the profiles of various quantities,
e.g. the density profile with $1/r^3$, as well as the shock radius
obtained here are in good agreement with the results of detailed simulations
for relatively low luminosities of
$\lesssim 6 \times 10^{52}{\rm ergs}$ ${\rm s}^{-1}$.
For higher luminosities, however, our results differ from
those of the simulations, suggesting that the latter are far
from the critical lines for shock revival.

\begin{figure}
\plotone{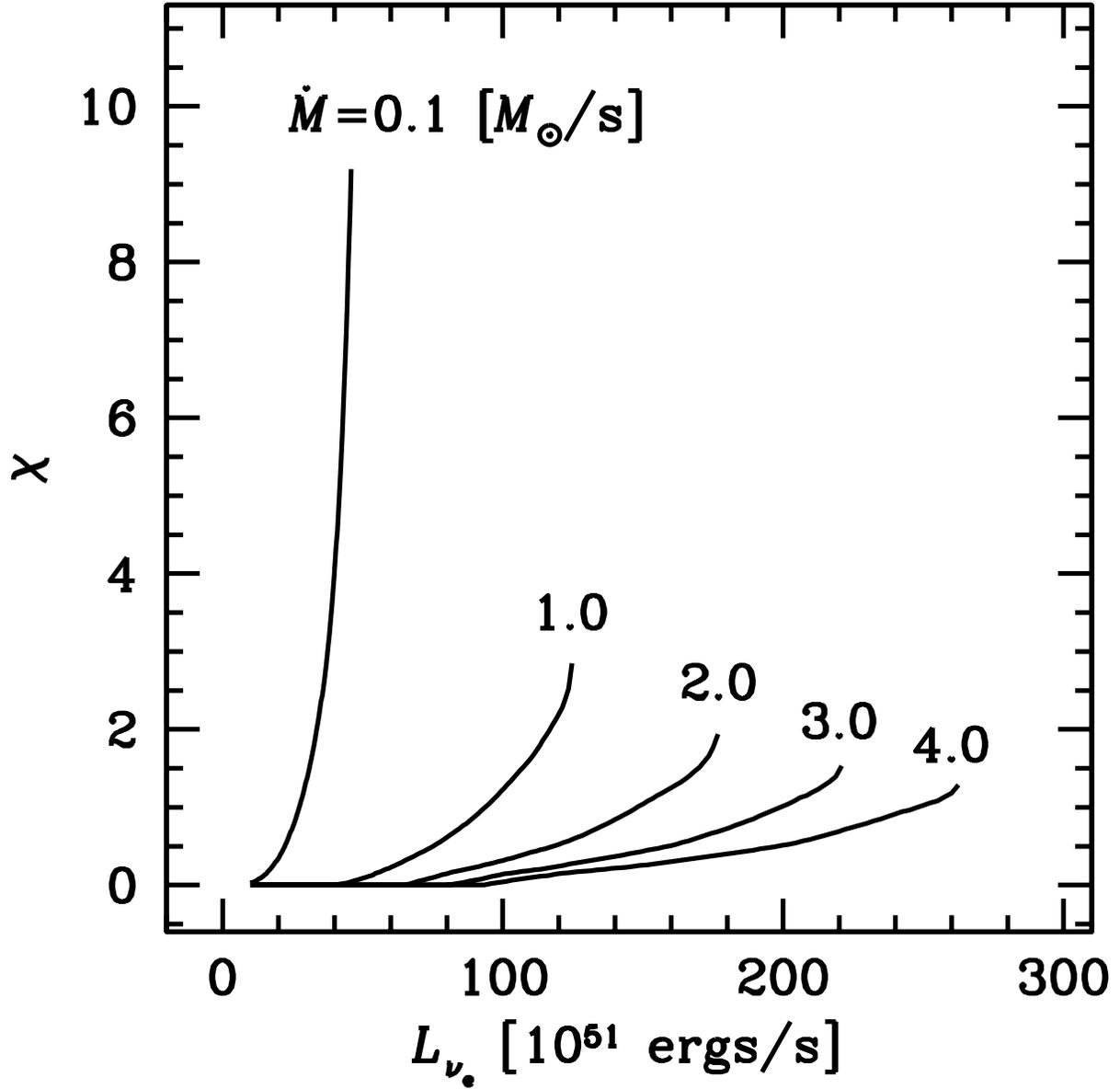}
\caption{The parameter $\chi$
for $\dot{M}=0.1, 1.0, 2.0, 3.0,
4.0$ ${M}_{\odot}$ ${\rm s}^{-1}$ from left to right.
The self-gravity is taken into account for these models.
\label{fig8}}
\end{figure}

In this paper, we have employed the classical
criterion for the convective stability, which is,
rigorously speaking, applicable only to the configuration without advection.
Recently, \citet{fog05} gave the modified criterion that 
takes into account the advection.
They introduced the parameter $\chi$ defined as
\begin{equation}
\chi\equiv\int_{r_{\rm gain}}^{r_{\rm sh}}
\left|\frac{N}{u_r}\right|dr,
\label{eq26}
\end{equation}
where $N$ is the Brunt-V$\ddot{\rm a}$is$\ddot{\rm a}$l$\ddot{\rm a}$
frequency given as
\begin{equation}
N^2 =\left\{\frac{1}{p}\left(\frac{\partial p}{\partial S}\right)_{\rho,Y_e}
\frac{dS}{dr}+\frac{1}{p}\left(\frac{\partial p}{\partial Y_e}\right)_{\rho,S}
\frac{dY_e}{dr}
\right\}\frac{g}{\Gamma_1}.
\label{eq27}
\end{equation}
In the above expression, $\Gamma_1$ and $g$ are given by
\begin{equation}
\Gamma_1 =\left(\frac{\partial\ln p}{\partial\ln \rho}\right)_{S,Y_e},
\label{eq28}
\end{equation}
and
\begin{equation}
g=\frac{GM}{r^2}.
\label{eq29}
\end{equation}
They claimed that the convection occurs in the presence of advection
if the entropy-gradient is negative and $\chi$ is larger than $3$.
In order to see the consequence of this modification to our models, 
we have calculated the values of $\chi$ for the solutions without convection 
taken into account.
The values of $\chi$ for our models with self-gravity are shown in
figure \ref{fig8}.
We can see that $\chi$ is always smaller than $3$
for the mass accretion rates larger than $1.0 {M}_{\odot}{\rm s}^{-1}$ and
that even for the mass accretion rate of $0.1 {M}_{\odot}{\rm s}^{-1}$, $\chi$
is smaller than $3$ except for a narrow range of neutrino luminosity.
Thus, taken at face values, the criterion by \citet{fog05} predicts
that the convection does not occur for most of our models.
This seems to be at odds with the results of numerical simulations.
It is, we think, worth notice that the steady solutions
we have obtained here are qualitatively different
from the unperturbed states assumed in \citet{fog05}.
\citet{fog05} assumed simplified formulae for the neutrino heating and
equation of state while we employed more realistic ones and, as a result,
the post-shock flows are much complicated.
Hence we suspect that the critical value of $\chi$ may be lowered in this case.
Indeed, the detailed linear analysis using our steady solutions
suggests that the critical value of $\chi$ may be smaller than $3$
\citep{yam06}.
However, the present models employing the classical criterion will give the
upper limit for the effect of convection.

There are a couple of things that have not been considered
in this paper but will be important. 
For example, we have introduced the phenomenological energy- and
proton-number fluxes in the convection region,
but have neglected the kinetic energy of turbulence.
This may not be ignored for the large velocity fluctuations
as observed in the numerical simulations.
Taken into account, it might increase the critical luminosity
because the energy of down-flows will be sucked.
It is also pointed out that the non-spherical deformation of the shock front
by convection may be an important ingredient
that cannot be considered in the present frame work. 
Thus far we have paid attention only to the condition for the shock revival.
This is, however, a minimum requirement.
The explosion energy and the mass of neutron star left behind are
equally important issues. We are currently attempting
to estimate these values based on the models presented here. 
Our model is also being extended to accommodate rotation, magnetic field,
anisotropic neutrino irradiation as well as being utilized
for the linear analysis of the non-local instability \citep{fog02,blo03,ohn06},
whose results will be published in the near future.

\acknowledgments

This work is partially supported by the Grant-in-Aid for the 21st century
COE program "Holistic Research and Education Center for Physics of
Self-organizing Systems" of Waseda University and for Scientific Research
(14740166, 14079202) of the Ministry of Education, Science, Sports and
Culture of Japan.

\clearpage

\begin{thebibliography}{}
\bibitem[Blondin et al.(2003)]{blo03} Blondin, J. M.,
    Mezzacappa, A., \& DeMarino, C.  2003, \apj, 584, 971
\bibitem[Bruenn(1985)]{bru85} Bruenn, A.  1985, \apjs, 58, 771
\bibitem[Buras et al.(2003)]{bur03} Buras, R.,
    Rampp, M.,Janka, H. T., \& Kifonidis, K.  2003, \prl,
    90, 241101
\bibitem[Buras et al.(2006)] {bur06} Buras, R., Janka, H. T., Rampp, M.,
    \& Kifonidis, K. 2006, \aap, in press (astro-ph/0512189)
\bibitem[Burrows \& Goshy(1993)]{bur93} Burrows, A.,
    \& Goshy, J.  1993, \apjl, 416, L75
\bibitem[Burrows et al.(1995)]{bur95} Burrows, A., Hayes, J.,
    \& Fryxell, B. A.  1995, \apj, 450, 830
\bibitem[Foglizzo(2002)]{fog02} Foglizzo, T.
    2002, \aap, 392, 353
\bibitem[Foglizzo et al.(2006)]{fog05} Foglizzo, T., Scheck, L.,
    \& Janka, H.-Th. 2006, \apj, in press (astro-ph/0507636)
\bibitem[Herant et al.(1994)]{her94} Herant, M., Benz, W., Hix, W. R.,
    Fryer, C. L., \& Colgate, S.A. 1994, \apj, 435, 339
\bibitem[Janka \& M\"{u}ller(1996)]{jan96} Janka, H.-Th. and M\"{u}ller, E. 
    1996, \aap, 306, 167
\bibitem[Liebend\"orfer et al.(2001)]{lie01} Liebend\"orfer, M.,
    Mezzacappa, A., Thielemann, F. K., Messer, O. E. B.,
    Hix, W. R., \& Bruenn, S. W., 2001, \prd, 63, 103004
\bibitem[Liebend\"orfer et al.(2005)]{lie05} Liebend\"orfer, M.,
    Rampp, M., Janka, H. T., \& Mezzacappa, A.,
    2005, \apj, 620, 840
\bibitem[Mezzacappa et al.(1998)]{mez98} Mezzacappa, A., Calder, A. C.,
    Bruenn, S. W., Blondin, J. M., Guidry, M. W., Strayer, M. R.,
    \& Umar, A. S., 1998 \apj, 493, 848
\bibitem[Ohnishi et al.(2006)]{ohn06} Ohnishi, N., Kotake K., \& Yamada, S.
    2006, \apj, 641, 1018
\bibitem[Shapiro and Teukolsky(1983)]{sha83} Shapiro, S. L.,
    \& Teukolsky, S. A.  1983, in Black Holes, White Dwarfs,
    and Neutron Stars (New York: John Wiely \& Sons, Inc.)
\bibitem[Shen et al.(1998)]{she98} Shen, H., Toki, H., Oyamatsu, K.,
    \& Sumiyoshi, K.  1998, \nphysa, 637, 435
\bibitem[Sumiyoshi et al.(2005)]{sum05} Sumiyoshi, K., Yamada, S., 
    Suzuki, S., Shen, H., Chiba, S.,\& Toki, H.  2005, \apj, 629, 922
\bibitem[Thompson et al.(2003)]{tho03} Thompson, T. A.,
    Burrows, A. \& Pinto, P. A. 2003 \apj, 592, 434
\bibitem[Wilson(1985)]{wil85} Wilson, J. R.
    1985, in Numerical Astrophysics, ed. J. M. Centrella, J. M. LeBlanc,
    \& R. L. Bowers (Boston: Jones and Bartlett), 422
\bibitem[Yamasaki \& Yamada(2005)]{yam05} Yamasaki, T.,
    \& Yamada, S.  2005, \apj, 623,1000
\bibitem[Yamasaki \& Yamada(2006)]{yam06} Yamasaki, T.,
    \& Yamada, S. 2006, \apj, submitted, (astro-ph/0606581)

\end{thebibliography}
\end{document}